\documentclass[twocolumn,pra,showpacs,preprintnumbers,superscriptaddress,amsmath,amssymb,10pt,aps]{revtex4}

\usepackage{mathptmx}
\usepackage{subfigure}
\usepackage{psfrag,graphicx}
\usepackage{dcolumn}
\usepackage{amsmath,amssymb}
\usepackage{bm}
\usepackage{color}
\usepackage{latexsym}
\usepackage{epstopdf}
\usepackage{color}
\usepackage[english]{babel}
\usepackage{latexsym}
\usepackage{psfrag,graphicx}
\usepackage{subfigure}
\usepackage{amsmath}
\usepackage{amssymb}
\usepackage{amsfonts}
\usepackage{bm}
\usepackage{natbib}
\usepackage{epstopdf}
\usepackage{appendix}

\definecolor{mygrey}{gray}{0.35}
\definecolor{myblue}{rgb}{0.2,0.2,0.8}
\definecolor{myzard}{cmyk}{0,0,0.05,0}
\definecolor{mywhite}{rgb}{1,1,1}
\definecolor{mywhite}{rgb}{1,1,1}
\definecolor{myred}{rgb}{1,0.,0.3}

\usepackage[colorlinks=true,citecolor=myblue,linkcolor=myred]{hyperref}

\def\ba{\begin{align}}
\def\enda{\end{align}}
\def\bi{\begin{itemize}}
\def\ei{\end{itemize}}

\def\be{\begin{equation}}
\def\ee{\end{equation}}
\def\bea{\begin{eqnarray}}
\def\eea{\end{eqnarray}}
\def\bse{\begin{subequations}}
\def\ese{\end{subequations}}

\begin{document}

\title{Adiabatic Sensing Technique for Optimal Temperature Estimation using Trapped Ions}

\author{Aleksandrina V. Kirkova}
\affiliation{Department of Physics, St. Kliment Ohridski University of Sofia, James Bourchier 5 blvd, 1164 Sofia, Bulgaria}
\author{Weibin Li}
\affiliation{School of Physics and Astronomy, and Centre for the Mathematics and Theoretical Physics of Quantum Non-Equilibrium Systems, University of Nottingham, NG7 2RD, United Kingdom}
\author{Peter A. Ivanov}
\affiliation{Department of Physics, St. Kliment Ohridski University of Sofia, James Bourchier 5 blvd, 1164 Sofia, Bulgaria}

\begin{abstract}
We propose an adiabatic method for optimal phonon temperature estimation using trapped ions which can be operated beyond the Lamb-Dicke regime. The quantum sensing technique relies on a time-dependent red-sideband transition of phonon modes, described by the non-linear Jaynes-Cummings model in general. A unique feature of our sensing technique is that the relevant information of the phonon thermal distributions can be transferred to the collective spin-degree of freedom. We show that each of the thermal state probabilities is adiabatically mapped onto the respective collective spin-excitation configuration and thus the temperature estimation is carried out simply by performing a spin-dependent laser fluorescence measurement at the end of the adiabatic transition. We characterize the temperature uncertainty in terms of the Fisher information and show that the state projection measurement saturates the fundamental quantum Cram\'er-Rao bound for quantum oscillator at thermal equilibrium.
\end{abstract}

\maketitle

\section{Introduction}
Over the last few years the devolvement of high-precision temperature sensing techniques has attracted considerable interest due to the broad and important applications ranging from medicine and biology \cite{Kinkert2009} to quantum information processing and quantum thermodynamics \cite{Mehboudi2019,Pasquale2016,Gemmer2004}. The quantum thermometer in generally consists of a system called probe which is brought into thermal equilibrium with a sample of interest. Various quantum optical systems can be used as temperature probes including for example quantum dots \cite{Seilmeier2014,Haupt2014,Sabin2014}, color centers in nanodiamonds \cite{Neumann2013,Kucsko2013,Toyli2013}, micromechanical resonators \cite{Brunelli2011,Brunelli2012} and trapped ions \cite{Meekhof1996,Robnagel2015,Gebert2016,Wan2015,Levy2020}.
An accurate strategy for temperature determination can be executed by measuring the populations in the energy basis of the quantum probe system \cite{Paris2016,Marzolino2013,Correa2015,Campbell2017,Campbell2018}. Indeed, it turns out that this strategy is \emph{optimal} with smallest temperature statistical uncertainty which saturates the fundamental Cram\'er-Rao bound for temperature estimation of any equilibrium system. However, the energy measurements are in general challenging as in case of a probe consisting of a quantum harmonic oscillator, where the number of basis states is typically large at thermal equilibrium, which limits the achievable temperature precision. Alternative approach is to use additional ancillary qubits to couple coherently with the probe. Then the information of the temperature is transferred to the qubit states which can be read-out with high-efficiency at the end of the interaction \cite{Brunelli2011,Brunelli2012,Ivanov2019}. Although this strategy is experimentally more convenient the statistical uncertainty of the temperature determination is usually higher than the optimal one given by the fundamental quantum Cram\'er-Rao bound.

In this work we propose an optimal adiabatic method for phonon temperature detection using trapped ions. Our technique relies on a global laser radiation which couples the internal spin states of ions to the vibrational mode via a red-sideband transition. This collective interaction is described by a non-linear Jaynes-Cummings type model in general. We show that by engineering time-dependent detuning and spin-motion coupling one can adiabatically transfer the relevant temperature information encoded in phonon distributions of vibrations onto the collective spin-excitation. Such a time-dependent control of the spin-phonon interaction has been extensively studied in creating of entangled spin and motion states \cite{Linington2008,Linington2008_1,Hume2009,Toyoda2011}. Here we show that each of the Fock states of the harmonic oscillator is adiabatically mapped on respective spin-excitation configuration. Thus the temperature determination is carried out by performing projection measurement of the spin populations at the end of the adiabatic transition. We show that our adiabatic sensing technique can be operated in and beyond the Lamb-Dicke limit and therefore is suitable for measuring a broad range of temperatures including a low temperature limit with mean thermal phonon excitations $\bar{n}\ll1$ as well as the high temperature regime with $\bar{n}\gg 1$. We quantify the sensitivity of the temperature estimation using classical Fisher information. We show that the projection measurements in the original spin basis lead to an \emph{equality} between the classical and quantum Fisher information for quantum harmonic oscillator at thermal equilibrium. Therefore, our quantum thermometry is optimal in the sense that the uncertainty of the temperature estimation is bounded by the quantum Cram\'er-Rao inequality. Moreover, we show that our adiabatic motion sensing technique can be applied for the detection of various other quantum states such as coherent and squeezed motion states. In particular, we discuss the detection of the phase of the coherent cat state via state-projective measurements which can be used for ultra sensitive force measurement with Heisenberg limit precision \cite{Maiwald2009,Munro2002}.

The paper is organized as follows: In Sec. \ref{background} we provide the general theoretical framework on the sensitivity of the temperature estimation. In Sec. \ref{realization} we discuss the physical realization of the adiabatic temperature estimation technique using trapped ions. The adiabatic method relies on a time-dependent red-sideband interaction which transfers the relevant temperature information onto the collective spin states. We show that the state projection measurements in the original basis leads to equality between the classical and quantum Fisher information and thus the temperature uncertainty is bounded by the quantum Cram\'er-Rao inequality. In Sec. \ref{imperfections} we investigate effects due to the physical imperfections on the sensitivity of our adiabatic quantum thermometer. In Sec. \ref{cat} we discuss application of the sensing technique for measuring the relative phase of the coherent cat state. We show that the phase can be determined by performing spin projective measurement with Heisenberg limit precision. Finally, the conclusions are presented in Sec. \ref{conclusions}.

\section{Principe of a Quantum Thermometry}\label{background}

We begin by considering a probe system which is represented by a simple quantum harmonic oscillator with Hamiltonian $\hat{H}=\hbar\omega\hat{a}^{\dag}\hat{a}$, where $\hat{a}^{\dag}$ and $\hat{a}$ are the creation and annihilation operators of bosonic excitation with frequency $\omega$. We assume that the harmonic oscillator is prepared at thermal equilibrium and is described by a canonical Gibbs state with density matrix $\hat{\rho}_{T}=e^{-\beta\hat{H}}/Z=\sum_{n=0}^{\infty}p_{n}|n\rangle\langle n|$. Here $|n\rangle$ is the $n$th Fock state of the harmonic oscillator with eigenenergy $E_{n}=n\hbar\omega$, $p_{n}=Z^{-1}e^{-\beta E_{n}}$ are the corresponding thermal state probabilities, $Z={\rm Tr}(e^{-\beta\hat{H}})$ the associated partition function, $\beta=1/k_{\rm B}T$ with $k_{\rm B}$ being the Boltzmann constant and $T$ is the temperature, the parameter we wish to estimate. Since, the temperature is not a direct observable its value can be extracted only by performing suitable measurements of other experimentally accessible observable. For this goal, consider a discrete set of measurements defined in terms of its corresponding positive-operator valued measure (POVM) $\{\hat{\Pi}_{n}\}$, with $\sum_{n}\hat{\Pi}_{n}=\mathbb{I}$. The corresponding classical Fisher information which quantifies the amount of information on the temperature of the system is given by \cite{Paris2009}
\begin{equation}
F_{\rm C}(T)=\sum_{n}\frac{\left(\partial_{T}P_{n}(T)\right)^{2}}{P_{n}(T)},\label{CFI}
\end{equation}
where $P_{n}(T)={\rm Tr}(\hat{\Pi}_{n}\hat{\rho}_{T})$ is the probability to get outcome $n$ from the performed measurement. Furthermore, the variance $\delta T$ of the temperature estimator is bounded by the Cram\'er-Rao inequality
\begin{equation}
\delta T\ge \frac{1}{\sqrt{\nu F_{\rm C}(T)}},
\end{equation}
where $\nu$ is the experimental repetitions.

The optimal strategy to measure the value of the temperature is however associated with a privileged observable which maximize the classical Fisher information and thus allows to determine the temperature with ultimate precision. Indeed, it is possible to show that the classical Fisher information is upper bounded by $F_{\rm C}(T)\le F_{\rm Q}(T)$, where $F_{\rm Q}(T)={\rm Tr}(\hat{\rho}_{T}\hat{L}^{2})$ is the quantum Fisher information. Here $\hat{L}$ is the symmetrical logarithmic derivative (SLD) operator, which satisfies the operator equation $\partial_{T}\hat{\rho}_{T}=(\hat{\rho}_{T}\hat{L}+\hat{L}\hat{\rho}_{T})/2$. Thus, the ultimate achievable precision of the temperature determination, optimized over all possible measurements is quantified by the quantum Cram\'er-Rao bound
\begin{equation}
\delta T\ge \frac{1}{\sqrt{\nu F_{\rm Q}(T)}}.\label{QRB}
\end{equation}
The eigenstates of the SLD operator $\hat{L}$ define the optimal measurement basis in which the quantum Cram\'er-Rao bound can be saturated. It is straightforward to show that for a Gibbs state with $\hat{\rho}_{T}$ the SLD operator can be written as $\hat{L}=\sum_{n}\{(E_{n}-\langle \hat{H}\rangle)/T^{2}\}|n \rangle\langle n|$, where $\langle \hat{H}\rangle={\rm Tr}(\hat{H}\hat{\rho}_{T})$ is the average energy \cite{Paris2016}. The result emphasizes that the optimal temperature measurement is achieved in the Fock basis $|n\rangle$ of the harmonic oscillator, e.g., by measuring the probabilities $p_{n}$. Finally, the QFI for the harmonic oscillator at thermal equilibrium can be written as
\begin{equation}
F_{\rm Q}(T)=\frac{\hbar^{2}\omega^{2}}{4k_{\rm B}^{2}T^{4}}{\rm csch}^{2}\left(\frac{\hbar\omega}{2k_{\rm B}T}\right).\label{QFI}
\end{equation}
A question that arises is whether it is possible to saturate the fundamental quantum Cram\'er-Rao bound by performing different set of discrete measurements rather than measurements of the thermal state probabilities. For this goal we consider an auxiliary quantum system of $N$ spin-$1/2$ particles which interacts coherently with the quantum harmonic oscillator. Using time-dependent unitary evolution one can map the information of the temperature into the respective spin state populations. We show that performing state projection measurements one can saturate the fundamental quantum Cram\'er-Rao bound and thus determine the temperature with the ultimate precision given by Eq. (\ref{QRB}).
\section{Ion Trap Realization of Quantum Thermometry}\label{realization}

We discuss in the following the ion trap based quantum thermometer which is able to perform an optimal measurement of the phonon temperature by detecting the ions' spin populations. We consider a linear ion crystal of $N$ ions confined in a Paul trap along the $z$ axis with trap frequencies $\omega_{\chi}$ ($\chi=x,y,z$). We assume that the transverse frequencies are much larger than the axial trap frequency $\omega_{x,y}\gg\omega_{z}$ which leads to the formation of a linear ion crystal where the ions occupy equilibrium positions $z_{k}^{0}$ along the trap axis. The position operator of the $l$th ion can be expressed as $\hat{r}_{l}=\delta \hat{r}_{x,l}\vec{e}_{x}+\delta \hat{r}_{y,l}\vec{e}_{y}+(z_{l}^{0}+\delta \hat{r}_{z,l})\vec{e}_{z}$, where $\delta \hat{r}_{\chi,l}$ are the displacement operators around the ion's equilibrium positions, which can be written in terms of collective phonon modes as $\delta \hat{r}_{\chi,l}=\sum_{k=1}^{N}M^{\chi}_{l,k}\sqrt{\frac{\hbar}{2m\omega_{\chi,k}}}(\hat{a}^{\dag}_{\chi,k}+\hat{a}_{\chi,k})$ \cite{James1998}. Here $\hat{a}^{\dag}_{\chi,k}$ and $\hat{a}_{\chi,k}$ are the creation and annihilation operators of the collective phonons with frequency $\omega_{\chi,k}$ along the spatial direction $\chi$. The element $M^{\chi}_{l,k}$ is the amplitude of the normal mode $k$ on ion $l$. We assume that each ion has two metastable internal levels $\left|\downarrow\right\rangle$ and $\left|\uparrow\right\rangle$ with a transition frequency $\omega_{0}$. Then, the interaction-free Hamiltonian describing the linear ion crystal is given by
\begin{equation}
\hat{H}_{0}=\hbar\omega_{0}\hat{S}_{z}+\hbar\sum_{k=1}^{N}\sum_{\chi=x,y,z}\omega_{\chi,k}\hat{a}^{\dag}_{\chi,k}\hat{a}_{\chi,k},\label{H0}
\end{equation}
where $\hat{S}_{z}=\frac{1}{2}\sum_{l=1}^{N}\sigma^{z}_{l}$ and $\hat{S}^{+}=\sum_{l=1}^{N}\sigma^{+}_{l}$ ($\hat{S}^{-}=(\hat{S}^{+})^{\dag}$) are the collective spin operators with $\sigma_{l}^{z}$ being the Pauli operator for the $l$th spin and respectively $\sigma^{+}_{l}=\left|\uparrow_{l}\right\rangle\left\langle\downarrow_{l}\right|$ is the spin raising operator.
\begin{figure}
\includegraphics[width=0.45\textwidth]{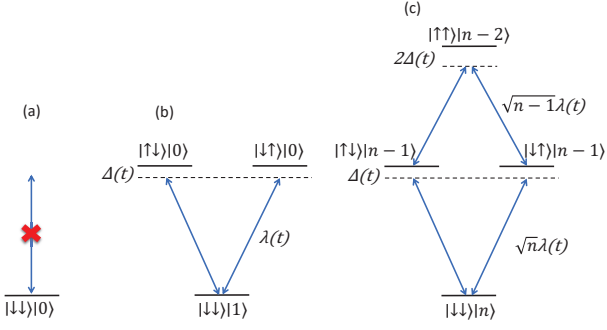}
\caption{(Color online) Linkage pattern of the collective states of a string of two ions driven by red-sideband laser. Spins are initially prepared in their electronic ground state and the vibration center-of-mass mode is in thermal states. a) The state $\left|\downarrow\downarrow\right\rangle\left|0\right\rangle$ is not affected by the collective red-sideband interactions. b) and c) The states $\left|\downarrow\downarrow\right\rangle\left|1\right\rangle$ and $\left|\downarrow\downarrow\right\rangle\left|n\right\rangle$ ($n>1$) are coupled to the manifolds with the same number of total excitations.}
\label{fig1}
\end{figure}

After performing a Doppler cooling of the linear ion crystal each collective vibrational mode is in a thermal state of motion with mean thermal phonon excitation $\bar{n}_{\chi,k}$. Since the oscillations of the ions in all three directions are decoupled one can determine the temperature of each vibrational mode independently \cite{Meekhof1996}. For concreteness we consider the temperature estimation of the collective center-of-mass mode along the spatial transverse direction $x$. This mode has the highest vibrational frequency $\omega_{x,1}=\omega_{x}$ in which the ions oscillate in phase with equal amplitude. The total Hilbert space is spanned by the basis $\{|S,m\rangle\otimes|n\rangle\}$ where $|n\rangle$ is the Fock state of the center-of-mass vibrational mode with $n$ phonons. The states $|S,m\rangle$ are the eigenvectors of the two commuting operators $\hat{S}^{2}|S,m\rangle=S(S+1)|S,m\rangle$ and $\hat{S}_{z}|S,m\rangle=m|S,m\rangle$ ($m=-S,\ldots,S$) with $S=\frac{N}{2}$. In the computational basis the state $\left|D_{l}\right\rangle=|S,-S+l\rangle$ with $l$ spin excitations ($l=0,1,\ldots, 2S$) can be expressed as
\begin{equation}
\left|D_{l}\right\rangle=\sqrt{\frac{l!(2S-l)!}{2S!}}\sum_{x}P_{x}\left|\uparrow_{1}\ldots\uparrow_{l}\downarrow_{l+1}\ldots\downarrow_{N}\right\rangle,
\end{equation}
where the sum subscript $x$ runs over all distinct permutations $P_{x}$ of the ions' internal states with $l$ spins in excited state $\left|\uparrow\right\rangle$ and respectively $N-l$ in the ground state $\left|\downarrow\right\rangle$.

In order to create a coupling between the collective vibrations and the ion spin states we assume that the linear ion crystal is globally addressed by laser field with laser wave vector $\vec{k}$ pointing along the $x$ direction ($|\vec{k}|=k_{x}$) and laser frequency $\omega_{\rm L}(t)=\omega_{0}-\omega_{x}+\Delta(t)$ tuned near the center-of-mass red-sideband resonance with time-dependent detuning $\Delta(t)$ ($\omega_{x}\gg\Delta(t)$). After performing an optical rotating-wave approximation, the interaction Hamiltonian becomes \cite{Wineland1998,Haffner2008,Schneider2012}
\begin{eqnarray}
\hat{H}_{I}(t)&=&\hbar\Omega(t)\sum_{l=1}^{N}\{\sigma^{+}_{l}e^{i(\sum_{k=1}^{N}\eta^{x}_{l,k}(\hat{a}^{\dag}_{x,k}e^{i\omega_{x,k}t}+\hat{a}_{x,k}e^{-i\omega_{x,k}t})}\notag\\
&&\times e^{i(\omega_{x}t-\int_{t_{i}}^{t}\Delta(\tau)d\tau)}+{\rm h.c.}\},\label{HI}
\end{eqnarray}
where $\Omega(t)$ is the time-dependent Rabi frequency and $\eta^{x}_{l,k}=k_{x}\sqrt{\frac{\hbar}{2m\omega_{x,k}}}M^{x}_{l,k}$ is the Lamb-Dicke parameter. Moreover, since the laser field frequency is close to the red-sideband resonance of the center-of-mass mode one can perform vibrational rotating-wave approximation, in which the contribution of the other spectator phonon modes is neglected. Transforming the Hamiltonian in the rotating-frame with respect to $\hat{U}_{\rm R}=e^{i\int_{t_{i}}^{t}\Delta(\tau)d\tau \hat{S}_{z}}$ such that $\hat{H}_{\rm JC}(t)=\hat{U}_{R}^{\dag}\hat{H}_{I}(t)\hat{U}_{R}-i\hbar \hat{U}_{R}^{\dag}\partial_{t}\hat{U}_{R}$, we arrive to
\begin{equation}
\hat{H}_{\rm nJC}(t)=\hbar \Delta(t)\hat{S}_{z}+\hbar \lambda(t)(\hat{S}^{+}\hat{F}(\hat{n})\hat{a}+\hat{S}^{-}\hat{a}^{\dag}\hat{F}(\hat{n})),\label{nHJC}
\end{equation}
\begin{figure}
\includegraphics[width=0.45\textwidth]{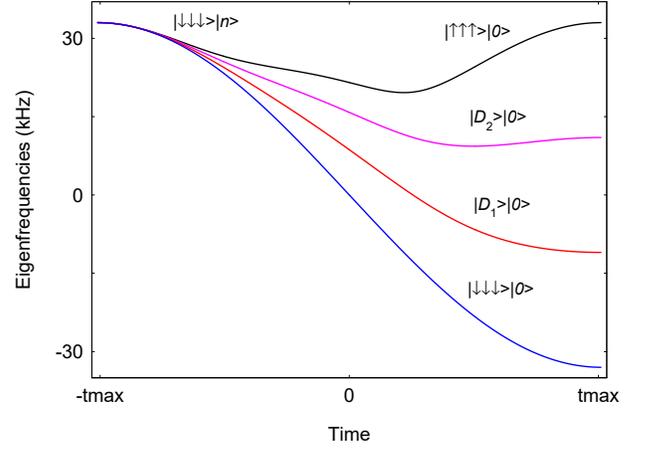}
\caption{(Color online) Lowest eigenfrequencies of Hamiltonian (\ref{HJC}) for three spins and for different phonon number $n$ as a function of time. We assume time-dependent detuning and spin-phonon coupling are given by Eq. (\ref{couplings}). In the adiabatic limit each of the initial states $|\psi_{n}(t_{i})\rangle=\left|\downarrow\downarrow\downarrow\right\rangle\left|n\right\rangle$ ($n=0,1,2,3$) is transformed into $|\psi_{n}(t_{i})\rangle\rightarrow|D_{n}\rangle|0\rangle$. }
\label{fig2}
\end{figure}where $\lambda(t)=\Omega(t)\eta^{x}_{l,1}$ is the time-dependent spin-phonon coupling and $\eta^{x}_{l,1}=\eta$ being the Lamb-Dicke parameter for the center-of-mass vibrational mode and $\hat{a}^{\dag}$ and $\hat{a}$ are respectively the phonon creation and annihilation operators corresponding to an oscillator with frequency $\omega_{x}$. The Hamiltonian (\ref{nHJC}) describes the non-linear Jaynes-Cummings (nJC) model, where the non-linear operator function can be expressed as \cite{Vogel1995}
\begin{equation}
\hat{F}(\hat{n})=e^{-\eta^{2}/2}\sum_{n=0}^{\infty}\frac{(-\eta^{2})^{n}}{n!(n+1)!}\hat{a}^{\dag n}\hat{a}^{n}.\label{F}
\end{equation}
Assuming the Lamb-Dicke limit $\eta\langle(\hat{a}^{\dag}+\hat{a})^{2}\rangle^{1/2}\ll 1$ in which the amplitudes of oscillations of the ions around their equilibrium positions are small compared to optical wavelength one can approximate the Hamiltonian (\ref{nHJC}) to
\begin{equation}
\hat{H}_{\rm JC}(t)=\hbar \Delta(t)\hat{S}_{z}+\hbar \lambda(t)(\hat{S}^{+}\hat{a}+\hat{S}^{-}\hat{a}^{\dag}),\label{HJC}
\end{equation}
which describes the linear Jaynes-Cummings (JC) model.
We note that the Lamb-Dicke approximation is justified for low temperatures and small $\eta\ll1$. However, with increasing temperature one would need to consider the nJC Hamiltonian (\ref{nHJC}) as the effect of the non-linear term (\ref{F}) becomes significant.
\begin{figure}
\includegraphics[width=0.45\textwidth]{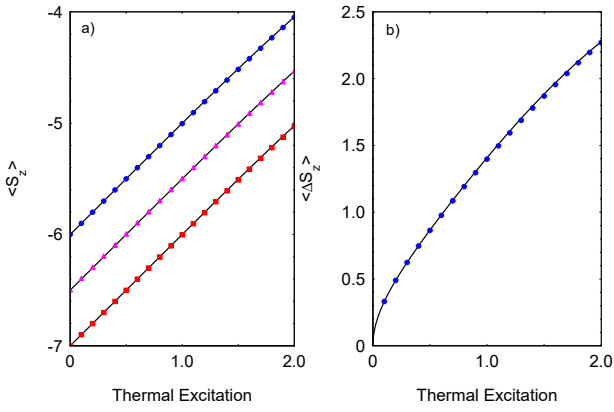}
\caption{(Color online) a) Average $\langle \hat{S}_{z}\rangle$ at $t_{\rm max}$ as a function of the thermal phonon excitation. We compare the result derived from the Hamiltonian $\hat{H}_{\rm JC}$ with the analytical solution (\ref{Sz}) (solid line) for $S=6$ (blue circles), $S=13/2$ (purple triangles) and $S=7$ (red squares). The other parameters are set to $\lambda_{0}/2\pi=5$ kHz, $\Delta_{0}/2\pi=22$ kHz, and $\gamma/2\pi=5.5$ kHz. b) The variance $\Delta \hat{S}_{z}$ at $t_{\rm max}$ for $S=6$. The blue circles are the exact solution and the solid line is the analytical expression (\ref{varS}).}
\label{fig3}
\end{figure}

Since the collective spin excitation can be created (annihilated) by absorption (emission) of collective center-of-mass phonon, the linear as well as the non-linear Jaynes-Cummings Hamiltonian commutes with the operator of the total number of excitations defined by $\hat{N}=\hat{S}_{z}+\hat{a}^{\dag}\hat{a}$. Consequently, the Hilbert space is decomposed into the subspaces with well defined total number of excitations $N=n_{\rm s}+n$ with $n_{\rm s}=0,1,\ldots,2S$ being the number of spin excitations.

\subsection{Temperature sensing protocol}

The temperature estimation scheme begins by preparing the system initially in the product state $\hat{\rho}_{i}=\hat{\rho}_{\rm spin}\otimes\hat{\rho}_{\rm th}$ where $\hat{\rho}_{\rm th}=\sum_{n=0}^{\infty}p_{n}\left|n\right\rangle\left\langle n\right|$ is the thermal state density operator for the center-of-mass mode with $p_{k}=\frac{\bar{n}^{k}}{(1+\bar{n})^{k+1}}$ and $\bar{n}=(e^{\beta\hbar\omega_{x}}-1)^{-1}$ being the average number of thermal excitations. We assume that the spins are initially polarized along the $z$-direction in a pure state with density matrix $\hat{\rho}_{\rm spin}=\left|D_{0}\right\rangle\left\langle D_{0}\right|$. Therefore, the initial total number of excitations is determined by the number of center-of-mass phonons $n$, namely $N=n$ ($n=0,1,2,\ldots$). Then the system evolves according the time-dependent red-sideband interaction such that the relevant temperature information is distributed over and stored in the collective spin degrees-of-freedom. In Fig. \ref{fig1} the linkage pattern of the collective states of linear crystal of two ions is shown where for concreteness we assume linear JC interaction described by Hamiltonian (\ref{HJC}). As it can be seen a collective spin excitation can be only created by the annihilation of center-of-mass phonon and vice versa. Thus the motional ground state is not affected by the red-sideband interaction, while states with $n>0$ phonons are coupled to the manifolds with the same number of total excitations. Since we deal with thermal motional states each of these three independent transitions is realized with probability $p_{n}$.
\subsection{Adiabatic Transition}
\begin{figure}
\includegraphics[width=0.45\textwidth]{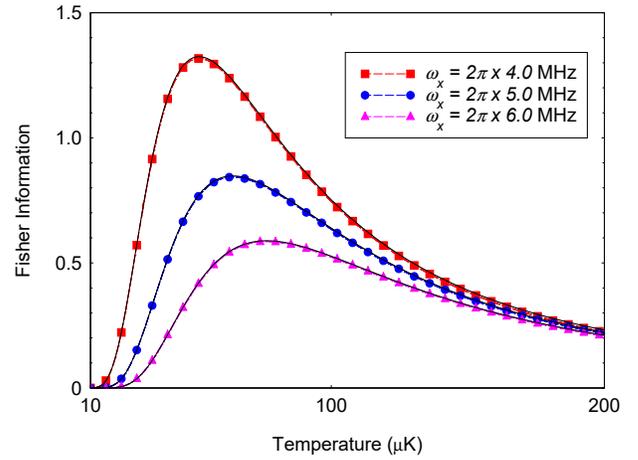}
\caption{(Color online) Classical Fisher information for the observables $P_{s_{1},\ldots,s_{N}}$ at $t_{\rm max}$ as a function of the temperature for ion chain with four ions. The numerical result for different transverse trap frequencies $\omega_{x}$ is compared with the QFI (\ref{QFI}) (solid lines). The other parameters are set to $\lambda_{0}/2\pi=5$ kHz, $\Delta_{0}/2\pi=25$ kHz, and $\gamma/2\pi=5.5$ kHz.  }
\label{fig4}
\end{figure}
Our goal is to determine the probabilities $p_{n}$ to observe a Fock state $\left|n\right\rangle$ by execute a projection spin-dependent measurements. First, we emphasize that due to the off-resonant transitions the application of $\pi$ laser pulse is not capable to distinguish the probabilities $p_{n}$ by measuring the spin population \cite{Wineland1998,Haffner2008,Schneider2012}. For this reason we adopt the adiabatic technique for detecting $p_{n}$ which is shower in time but more robust with respect to parameter fluctuation. In Fig. \ref{fig2} we show the lowest eigenfrequencies of Hamiltonian (\ref{HJC}) for three spins and different phonon numbers $(n=0,1,2,3)$. Assume that at the initial moment the laser detuning is much higher than the spin-phonon coupling, $|\Delta(t_{i})|\gg\lambda(t_{i})$ and $\Delta(t_{i})<0$. Then the state vectors $\left|\psi_{n}(t_{i})\right\rangle=\left|D_{0}\right\rangle\left|n\right\rangle$ are an eigenstates of Hamiltonian (\ref{HJC}) such that $\hat{H}_{\rm JC}(t_{i})\left|\psi_{n}(t_{i})\right\rangle=-S\Delta(t_{i})\left|\psi_{n}(t_{i})\right\rangle$. Adiabatically varying the detuning $\Delta(t)$ such that we end up with $\Delta(t_{f})\gg\lambda(t_{f})$ and $\Delta(t_{f})>0$. In the adiabatic limit, the system remains in the same eigenstate of the Hamiltonian (\ref{HJC}) at all times. Since the total number of excitations is preserved the initial state $|\psi_{n}(t_{i})\rangle$ is adiabatically transformed into the final state $|\psi_{n}(t_{f})\rangle=|D_{n}\rangle|0\rangle$ where we assume $n\le 2S$. Since the maximal number of spin excitations is $n_{\rm s}=2S$ in which all spins are in the excited levels, the initial state $|\psi_{n}(t_{i})\rangle$ with $n>2S$ adiabatically evolves into $|\psi_{n}(t_{f})\rangle=|D_{2S}\rangle|n-2S\rangle$. Therefore, for a state with $N$ spins and thermal motion state this implies the following transition
\begin{equation}
\hat{\rho}_{i}\rightarrow\hat{\rho}_{f}=\sum_{l=0}^{2S}p_{l}|D_{l}\rangle\langle D_{l}|\otimes|0\rangle\langle0|+\hat{\rho}_{\rm res}.\label{rho}
\end{equation}
Hence, the maximally mixed thermal motion state is adiabatically transformed into the maximally mixed spin state in which one can observe state $|D_{l}\rangle$ with probability $p_{l}$. Finally, the residual density matrix in (\ref{rho}) is given by
\begin{equation}
\hat{\rho}_{\rm res}=|D_{2S}\rangle\langle D_{2S}|\otimes\sum_{n=2S+1}^{\infty}p_{n}|n-2S\rangle\langle n-2S|.\label{res}
\end{equation}

A convenient choice of the time-dependent detuning and spin-boson coupling, which can be used to drive the adiabatic transition, is
\begin{equation}
\Delta(t)=\Delta_{0}\sin\left(\frac{\gamma t}{2}\right),\quad \lambda(t)=\lambda_{0}\cos^{2}\left(\frac{\gamma t}{2}\right)\label{couplings},
\end{equation}
where $\Delta_{0}>0$, $\lambda_{0}>0$ and $\gamma$ is a characteristic parameter which controls the adiabaticity of the transition. The interaction time varies as $t\in[-t_{\rm max} ,t_{\rm max}]$ with $t_{\rm max}=\pi/\gamma$ which ensures that $|\Delta(-t_{\rm max})|\gg \lambda(-t_{\rm max})$ and respectively $\Delta(t_{\rm max})\gg \lambda(t_{\rm max})$.
\begin{figure}
\includegraphics[width=0.45\textwidth]{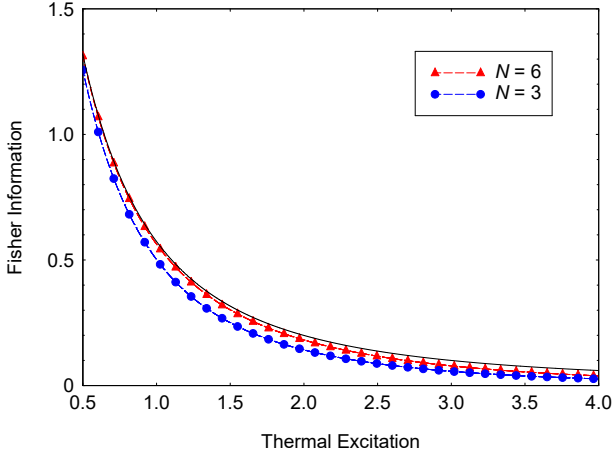}
\caption{(Color online) Classical Fisher information as a function of the thermal phonon excitation $\bar{n}$. The numerical result for $\omega_{x}/2\pi=6$ MHz and different number of ions is compared with the QFI (\ref{QFI}) (dashed lines).}
\label{fig5}
\end{figure}

In Fig. \ref{fig3}(a) we show the exact result for the average spin magnetization $\langle\hat{S}_{z}(t_{f})\rangle={\rm Tr}(\hat{\rho}_{f}\hat{S}_{z})$ compared with the analytical result given by
\begin{equation}
\langle \hat{S}_{z}(t_{f})\rangle=\bar{n}-S-\left(\frac{\bar{n}}{1+\bar{n}}\right)^{2S+1}(\bar{n}+S+1)\label{Sz},
\end{equation}
where very good agreement is observed. We see that the time-dependent red-sideband interaction rotates the initial spin magnetization which varies with the thermal phonon excitations and thus the observable $\langle \hat{S}_{z}(t_{f})\rangle$ can be used for detecting the temperature. Indeed, the shot-noise limited sensitivity in the temperature estimation from the measured signal $\langle \hat{S}_{z}(t_{f})\rangle$ is $\delta T^{2}=(\nu F_{S_{z}})^{-1}$ where $F_{S_{z}}=\frac{1}{\langle\Delta \hat{S}_{z}\rangle^{2}}\left(\frac{\partial\langle \hat{S}_{z}\rangle}{\partial T}\right)^{2}$ is the fidelity susceptibility \cite{Pezze2018}
and $\langle\Delta \hat{S}_{z}\rangle^{2}=\langle \hat{S}^{2}_{z}\rangle-\langle \hat{S}_{z}\rangle^{2}$ is the variance of $\hat{S}_{z}$. Using (\ref{rho}) it is straightforward to show that (see Fig. \ref{fig3}(b))
\begin{eqnarray}
\langle\Delta \hat{S}_{z}(t_{f})\rangle^{2}&=&\frac{\bar{n}}{(1+\bar{n})^{4S+2}}\{(1+\bar{n})^{4S+3}-\bar{n}^{4S+1}(1+S+\bar{n})^{2}\notag\\
&&-\bar{n}^{2S}(1+\bar{n})^{2S+1}[1+\bar{n}+S(4+3S+2\bar{n})]\}.\label{varS}
\end{eqnarray}
However, a more convenient approach for temperature estimation is to detect the spin populations $P_{s_{1},\ldots,s_{N}}={\rm Tr}(\hat{\rho}_{f}\hat{\Pi}_{s_{1},\ldots,s_{N}})$, where $\hat{\Pi}_{s_{1},\ldots,s_{N}}=|s_{1},\ldots,s_{N}\rangle\langle s_{N},\ldots,s_{1}|$ is the projection operator with $s_{l}=\uparrow_{l},\downarrow_{l}$. Indeed, the magnetization of each spin after the adiabatic transition can be measured by illuminating the ions with a global laser radiation and collecting the state-dependent fluorescence on a camera.
\begin{figure}
\includegraphics[width=0.45\textwidth]{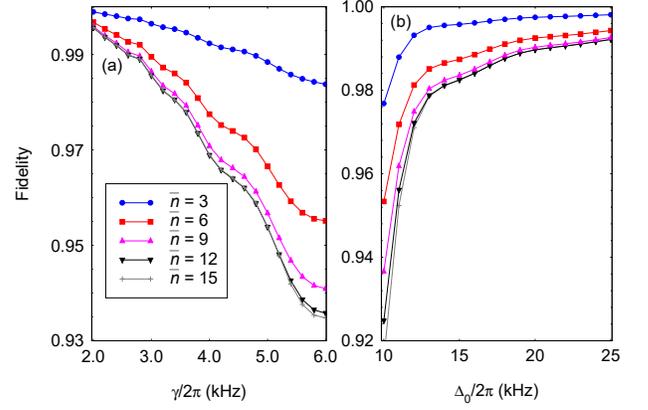}
\caption{(Color online) (a) Fidelity (\ref{fidelity}) at $t_{\rm max}$ for different characteristic rate $\gamma$. We integrate numerically the Liouville equation with Hamiltonian (\ref{nHJC}). The other parameters are set to $\lambda_{0}/2\pi=5$ kHz, $\Delta_{0}/2\pi=22$ kHz, $\eta=0.2$ and $N=6$. (b) The same but we vary the detuning $\Delta_{0}$ for $\gamma/2\pi=2.5$ kHz. }
\label{fig6}
\end{figure}

In Fig. \ref{fig4} we show the exact result for the classical Fisher information (\ref{CFI}) for the spin probabilities $P_{s_{1},\ldots,s_{N}}$ compared with the QFI (\ref{QFI}). We see that the classical Fisher information associated with the observables $P_{s_{1},\ldots,s_{N}}$ is equal to the quantum Fisher information (\ref{QFI}) for quantum harmonic oscillator at thermal equilibrium. Therefore, the detection of the orientation of each spin is optimal for the temperature estimation in the sense that the temperature uncertainty is bounded by the quantum Cram\'er-Rao bound (\ref{QRB}).
In Fig. \ref{fig5} is shown the numerical result for the classical Fisher information for different number of ions and high temperature. As the mean thermal phonon excitation increases the residual density matrix term $\hat{\rho}_{\rm res}$ (\ref{res}) limits temperature sensitivity.
Indeed, the probability to observe a collective state with all spins in the excited levels is not equal to $p_{2S}$ but other highly excited thermal phonon states with probabilities $p_{n}$ ($n>2S$) are also contributed, which spoil the optimal temperature estimation. However, as we can see from the Fig. \ref{fig5} the effect of the residual term can be suppressed by increasing the number of ions. Indeed, for higher number of ions the probability to observe all spins in the excited states after the adiabatic transition becomes negligibly small, so that the effect of the residual term $\hat{\rho}_{\rm res}$ can be suppressed which ultimately improves the temperature sensitivity.

In the following we examine the effect of the non-adiabatic transitions which limit the efficiency of the temperature determination. We discuss the red-sideband interaction beyond the Lamb-Dicke approximation by including the non-linear terms ($\ref{F}$), which becomes significant in the high temperature limit. Since the nJC Hamiltonian (\ref{nHJC}) preserves the total number of excitations the adiabatic transition (\ref{rho}) still holds. We show that the effect of the non-linear terms is merely to modify the adiabaticity of the transition.
\section{Physical Imperfections}\label{imperfections}
\begin{figure}
\includegraphics[width=0.45\textwidth]{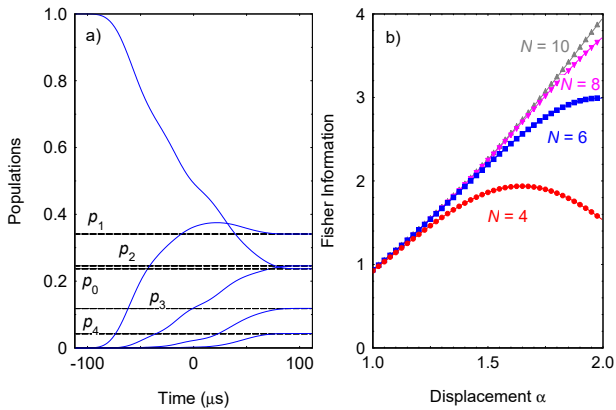}
\caption{(Color online) a) Collective spin populations as a function of time. We assume that the system is prepared in motion coherent state with density matrix operator $\hat{\rho}_{\alpha}=|\alpha\rangle\langle\alpha|$ with $\alpha=1.2$. The other parameters are set to $\lambda_{0}/2\pi=5$ kHz, $\Delta_{0}/2\pi=20$ kHz and $\gamma/2\pi=4.5$ kHz b) Classical Fisher information for the estimation very weak force $\epsilon$ for initial coherent cat state as a function of the displacement amplitude $\alpha$. The spin observables are measured at $t_{\rm max}$. We numerically integrate the Liouville equation with Hamiltonian (\ref{HJC}) for different number of ions. The other parameters are set to $\Delta_{0}/2\pi=22$ kHz and $\gamma/2\pi=2.2$ kHz.  }
\label{fig7}
\end{figure}
As a figure of merit for the efficiency of the adiabatic transition we use the fidelity between two density matrices defined by \cite{Gu2010}
\begin{equation}
F(\hat{\rho}_{f},\hat{\rho}(t))=\frac{{\rm Tr}(\hat{\rho}_{f}\hat{\rho}(t))}{\sqrt{{\rm Tr}(\hat{\rho}_{f}^{2}){\rm Tr}(\hat{\rho}(t)^{2})}}.\label{fidelity}
\end{equation}
Here $\hat{\rho}_{f}$ is the desired density matrix (\ref{rho}) and $\hat{\rho}(t)$ is the actual one. In Fig. \ref{fig6}(a) we show the numerical result for the fidelity (\ref{fidelity}) as a function of the controlling parameter $\gamma$ using the nJC Hamiltonian (\ref{nHJC}). As the temperature increases the Lamb-Dicke approximation is not fulfilled and thus one needs to include the high-order terms in the Lamb-Dicke expansion given by Eq. (\ref{F}). We observe that on one hand the non-adiabatic transitions become stronger for higher values of $\bar{n}$ and the fidelity decreases slightly when $\bar{n}$ increases toward high temperature limit. On the other hand the adiabaticity is improved for lower value of $\gamma$ and thus longer interaction time. For example, assuming the mean thermal phonon excitations $\bar{n}=15$ and $\gamma/2\pi=2.4$ kHz such that the total interaction time is $\tau=2 t_{\rm max}\approx 417$ $\mu$s, we estimate fidelity $F(\hat{\rho}_{f},\hat{\rho}(t_{\rm max}))>0.99$. Increasing the interaction time improves the fidelity until the random noise compromises the signal. For example, the electric fluctuations of the trap electrodes affect the motional phonon population during the adiabatic transition. Consider heating rate $\langle \dot{n}\rangle=1/t_{\rm dec}$, where $t_{\rm dec}$ is the characteristic decoherence time we require $t_{\rm dec}\gg \tau$. For heating rate of $0.1$ ${\rm ms}^{-1}$ \cite{Chiaverini2014}, which corresponds to typical heating rate in linear ion Pual traps and interaction time of order of $\tau\approx 0.4$ ms this condition is satisfied. Other possible source of errors are spontaneous spin flip from the excited state during the adiabatic transition and magnetic field fluctuations which cause spin dephasing. Usually the spontaneous decay of the upper level takes too long time of order of 1 s and thus it can be neglected. The coherence time is often limited by ambient magnetic field fluctuations which can be suppressed by using magnetic field insensitive transitions \cite{Lee2005}.

In Fig. \ref{fig6}(b) we show the fidelity as a function of the detuning $\Delta_{0}$ and for fixed $\gamma$. On one hand, as can be seen by increasing $\Delta_{0}$ the adiabaticity of the transition is improved which leads to higher fidelity. On the other hand in order to resolve the vibrational center-of-mass mode the energy splitting to the energetically nearest rocking mode with frequency $\omega_{\rm roc}=\sqrt{\omega_{x}^{2}-\omega_{z}^{2}}$ has to be sufficiently large compared to the spin-phonon coupling $\lambda_{0}$ and laser detuning $\Delta_{0}$, namely $\Delta_{\rm gap}\gg \lambda_{0},\Delta_{0}$ where $\Delta_{\rm gap}=\omega_{x}-\omega_{\rm roc}$. Increasing the number of ions however makes the vibrational modes closer, such that the laser addressability of the center-of-mass mode imposes a restriction on $N$. Moreover, for given aspect ratio $\omega_{z}/\omega_{x}$ there is a maximal number of ions for which the system undergoes structural phase transition to a zigzag configuration. Thus the energy gap scales with the number of ions as $\Delta_{\rm gap}/\omega_{x}\approx 0.6228 \ln(6N)/N^{2}$; see for more details \cite{Ivanov2013}. Consider $N=12$ and $\omega_{x}/2\pi=8$ MHz we find $\Delta_{\rm gap}/2\pi\approx148$ kHz. For $\gamma/2\pi=2.3$ kHz, $\bar{n}=6$ and $\Delta_{0}/2\pi=15$ kHz we estimate fidelity $F(\hat{\rho}_{f},\hat{\rho}(t_{\rm max}))>0.99$.

\section{Detection of the relative phase of the coherent cat state}\label{cat}

Let us extend the discussion by considering various initial motion states. In Fig. \ref{fig7}(a) we show the time evolution of the collective-spin states for initial coherent state. The adiabatic evolution drives the system into the final density matrix given by Eq. (\ref{rho}) where now the Fock state distribution is given by $p_{n}=e^{-|\alpha|^{2}}|\alpha|^{2n}/n!$. Thus, the relevant information of the magnitude of the displacement amplitude is mapped onto the collective spin excitations and thereby  it can be measured by detecting the spin populations at the end of the adiabatic transition.  Furthermore, our adiabatic technique can be applied also for detecting the relative phase of the coherent cat state. Consider a motional density matrix $\hat{\rho}_{\rm cat}=|\psi_{\rm cat}\rangle\langle\psi_{\rm cat}|$, where $|\psi_{\rm cat}\rangle=(|\alpha\rangle+\left|-\alpha\right\rangle)/\sqrt{2}$ is a coherent cat state ($\alpha\gg 1$) such that we have $\hat{\rho}_{i}=|D_{0}\rangle\langle D_{0}|\otimes\hat{\rho}_{\rm cat}$. We assume that a time-varying force is applied which is on resonance with the frequency of the center-of-mass mode. The effect of the force is to displace a small motion amplitude with $\hat{D}(\epsilon)=e^{i\epsilon(\hat{a}^{\dag}-\hat{a})}$ where $\epsilon$ is the parameter we wish to estimate. The information of $\epsilon$ ($\epsilon\ll1$) is imprinted in the relative phase of the coherent cat state, namely $|\psi_{\rm cat}\rangle\approx(e^{i\theta}|\alpha\rangle+e^{-i\theta}\left|-\alpha\right\rangle)/\sqrt{2}$, where $\theta=\alpha\epsilon$ \cite{Munro2002}. Then the system evolves according the time-dependent detuning $\Delta(t)$ and spin-phonon coupling $\lambda(t)$ (\ref{couplings}) such that at $t_{\rm max}$ the spin populations are measured. In Fig. \ref{fig7}(b) we show the exact result for the classical Fisher information for estimating $\epsilon$ as a function of the initial displacement amplitude $\alpha$ and for different number of ions. Crucially, using a coherent cat state, the precision in estimating $\epsilon$ grows quadratically with $\alpha$ which corresponds to a Heisenberg limit \cite{Munro2002}. In order to achieve such precision $\delta\epsilon^{2}\ge1/\alpha^{2}$ one needs to perform a state-dependent measurement on the spin states at the end of the adiabatic transition. As is shown in Fig. \ref{fig7}(b) increasing $\alpha$ results in more phonon states being populated which in turn requires the increase of the number of ions.

\section{Conclusions}\label{conclusions}.

We have proposed an efficient adiabatic method for temperature measurement with trapped ions which can be operated beyond the Lamb-Dicke limit. The technique is based on an adiabatic evolution which transfer the relevant phonon temperature information onto the spin populations which can be measured by state-dependent fluorescence at the end of the adiabatic transition with high efficiency. We have characterized the amount of temperature information which can be extracted for such a spin detection in terms of classical Fisher information. We have shown that the state-projection measurements lead to equality between the classical and quantum Fisher information for harmonic oscillators at thermal equilibrium. Thus the temperature is determined with ultimate precision given by the quantum Cram\'er-Rao bound.

Furthermore, we have discussed the application of our method for the detection of the relative phase of the coherent cat state. Such a phase can be generated by the application of very weak time-varying force which displaces the initial motional coherent cat state. We have shown that by executing a state projective measurement one can determine the unknown displacement with Heisenberg limit precision.

\section*{Acknowledgments}

A. V. K. and P. A. I. acknowledges support by the ERyQSenS project, Bulgarian Science Fund Grant No. DO02/3. W. L. acknowledges support from the EPSRC through grant No. EP/R04340X/1 via the QuantERA project “ERyQSenS”.


\end{document}